\documentclass[prl,twocolumn,aps,showpacs,nofootinbib,floatfix,groupedaddress]{revtex4-1}
\usepackage{amsmath}
\usepackage{amssymb}
\usepackage{graphicx}
\usepackage{verbatim}
\usepackage[colorlinks=true,linkcolor=blue,citecolor=blue,urlcolor=blue]{hyperref}
\usepackage{txfonts}

\begin{document}

\title{Cavity-mediated unconventional pairing in ultracold fermionic atoms}
\author{Frank Schlawin and Dieter Jaksch}
\address{Clarendon Laboratory, University of Oxford, Parks Road, Oxford OX1 3PU, United Kingdom}
\email{frank.schlawin@physics.ox.ac.uk}

\begin{abstract}

We investigate long-range pairing interactions between ultracold fermionic atoms confined in an optical lattice which are mediated by the coupling to a cavity. 
In the absence of other perturbations, we find three degenerate pairing symmetries for a two-dimensional square lattice.
By tuning a weak local atomic interaction via a Feshbach resonance or by tuning a weak magnetic field, the superfluid system can be driven from a topologically trivial $s$-wave to topologically ordered, chiral superfluids containing Majorana edge states. 
Our work points out a novel path towards the creation of exotic superfluid states by exploiting the competition between long-range and short-range interactions.

\end{abstract}

\maketitle

Ultracold fermionic gases form an ideal platform for a new generation of quantum technologies \cite{Rushton14, Keil16}, and for the quantum simulation of many-body phenomena \cite{Bloch08, Chin10} such as superfluidity \cite{Greiner03, Chin06}. Their key feature for these applications is that local (on-site) atomic interactions can be tuned very precisely using Feshbach resonances to explore, e.g, the crossover from BCS to BEC regimes \cite{Randeria14}, quantum simulation of the Fermi Hubbard model \cite{Stadler12}, and strongly correlated fermions in reduced dimensions \cite{Burger01, Chiofalo02, Stajic04, DePalo04}. Coupling ultracold atoms to optical cavities~\cite{Jaksch01, Baumann10, Goldbart11, Mekhov12, Mottl12, Ritsch13, Landig16, Leonard17,Kroeze18, Landini18} promises to extend this control to long-ranged interactions. 

This coupling can be employed to induce a self-ordering transition of the atomic gas which coincides with a superradiant transition of the cavity \cite{Keeling14, Chen14, Piazza14, Mivehvar17}. Artificial magnetic fields can be induced to create distinct topological phases~\cite{Pan15, Kollath16, Sheikhan16}. Most importantly, various types of cavity-mediated interactions could give rise to a plethora of many-body phases \cite{Guo12, Elliott15, Santiago15, Santiago16, Santiago16b, Camacho17, Colella18, Fan18, Mivehvar19}, including superfluid and charge density states, and even more exotic phases with no direct analog in condensed matter systems. 
All these developments render ultracold fermionic atoms natural candidates to explore exotic physics, such as topological phases \cite{Jotzu14, Mancini15, Zoller_review}, which would be more difficult to observe in condensed matter. 

Topological superfluids in particular constitute a highly interesting class of such states, as the zero-energy edge states feature Majorana fermions - exotic quasiparticles which are their own antiparticles and the elusive building blocks of future topological quantum computers \cite{Kallin16, Sato17}.
To date, numerous theoretical proposals to create such superfluids from ultracold fermions have been put forth. The most straightforward proposal, to induce pairing with $p$-wave Feshbach resonances \cite{Fedorov17, Wasseem17}, suffers from large particle loss rates. Theoretical proposals to circumvent this fundamental problem aim to combine conventional $s$-wave pairing with the design of complex single-particle Hamiltonians. Most prominently, Raman-mediated spin-orbit coupling can induce a similar effective atomic interaction \cite{Jiang11, Williams13, Midtgaard16, Wu16, Midtgaard17, Midtgaard18}, but to date cooling such a system to the superfluid phase transition has not been achieved~\cite{Zhai15}. Other proposals include spin-dependent optical lattices \cite{Liu14, Wang16, Isaev17}, tilted lattices \cite{Sato09}, or multi-species lattices \cite{Nascimbene13, Buehler14}. But to date neither of these proposals could be realised experimentally either, requiring delicately engineered Hamiltonians that are challenging to realize. 

Here we demonstrate how the competition between long-range cavity-mediated interactions and weak symmetry-breaking perturbations can be exploited to create a second-order topological $p+id$-wave state  - a novel class of topological state featuring Majorana fermions in corner states which to the best of our knowledge have not been observed to date.
These states arise among a wealth of other topological states that can be robustly created and fully controlled by engineering local atomic interactions. In particular, we find that in the absence of local interactions, three pairing symmetries - $s$-, $p$- and $d$-waves - are degenerate in a cubic lattice potential. The competition between the different phases leaves the system in a frustrated state, where symmetries can be spontaneously broken. 
Weak perturbations remove the degeneracy and push the system either into a topologically trivial $s$-wave, a first-order topological $p$-wave, or into the $p+i d$-state. 
The emergence of these exotic states relies only on the degeneracy of different pairing symmetries as a consequence of the long-range nature of the interactions, rather than the fine-tuning of a microscopic Hamiltonian. As such, we believe that these results are a generic consequence of the distinct length scales, and in this sense a robust feature of the long-range nature of the interaction induced by the cavity.
Hence, our results establish a close connection between unconventional pairing states and long-range interactions, which could be explored with current state-of-the-art technology.

\begin{figure}[t]
\centering
\includegraphics[width=0.3\textwidth]{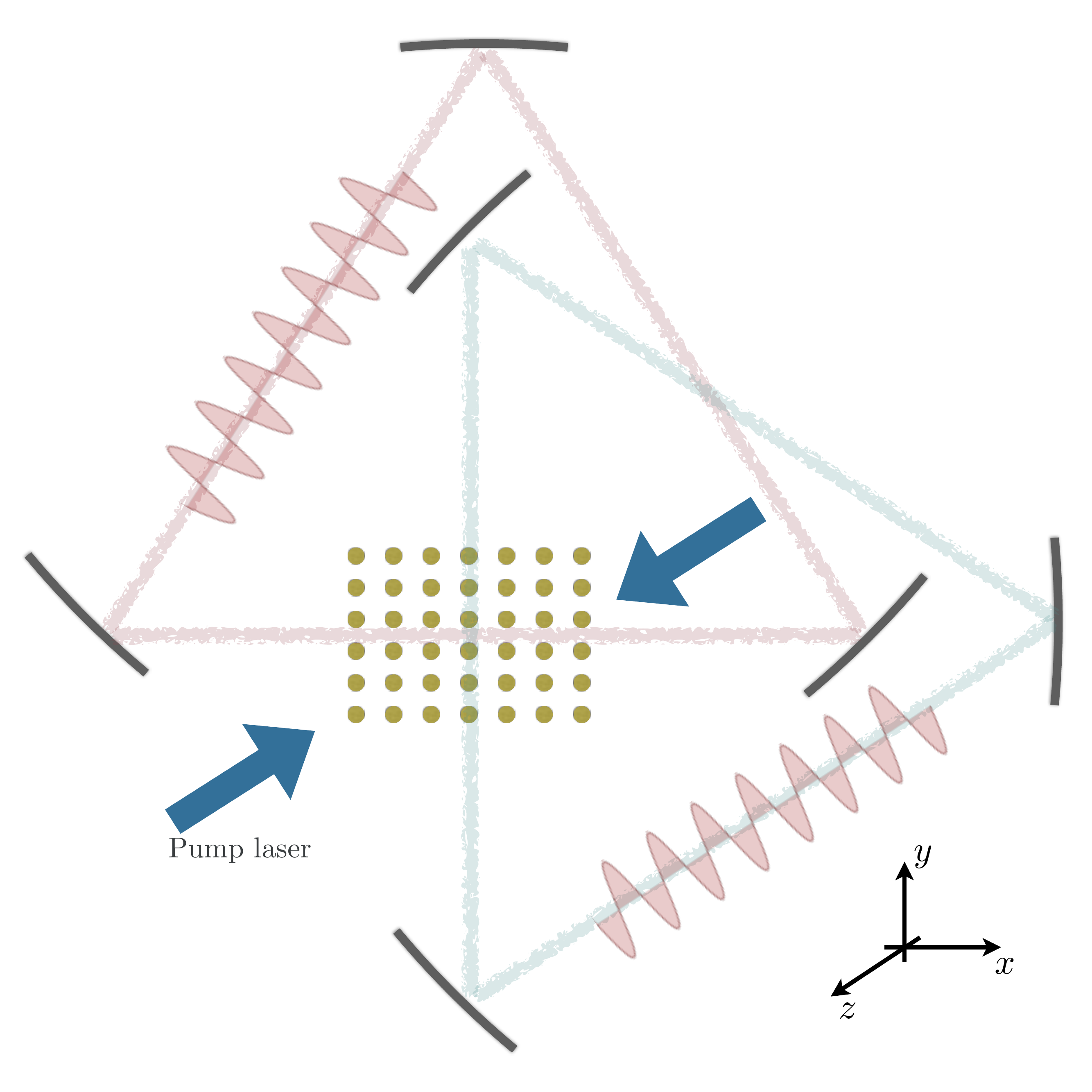}
\caption{
A fermionic atom gas is placed into a rectangular optical lattice, sketched by the yellow sites. The pump laser in combination with the two cavities induces long-range interactions between the atoms.
}
\label{fig.setup}
\end{figure}

\begin{figure*}[t]
\centering
\includegraphics[width=0.9\textwidth]{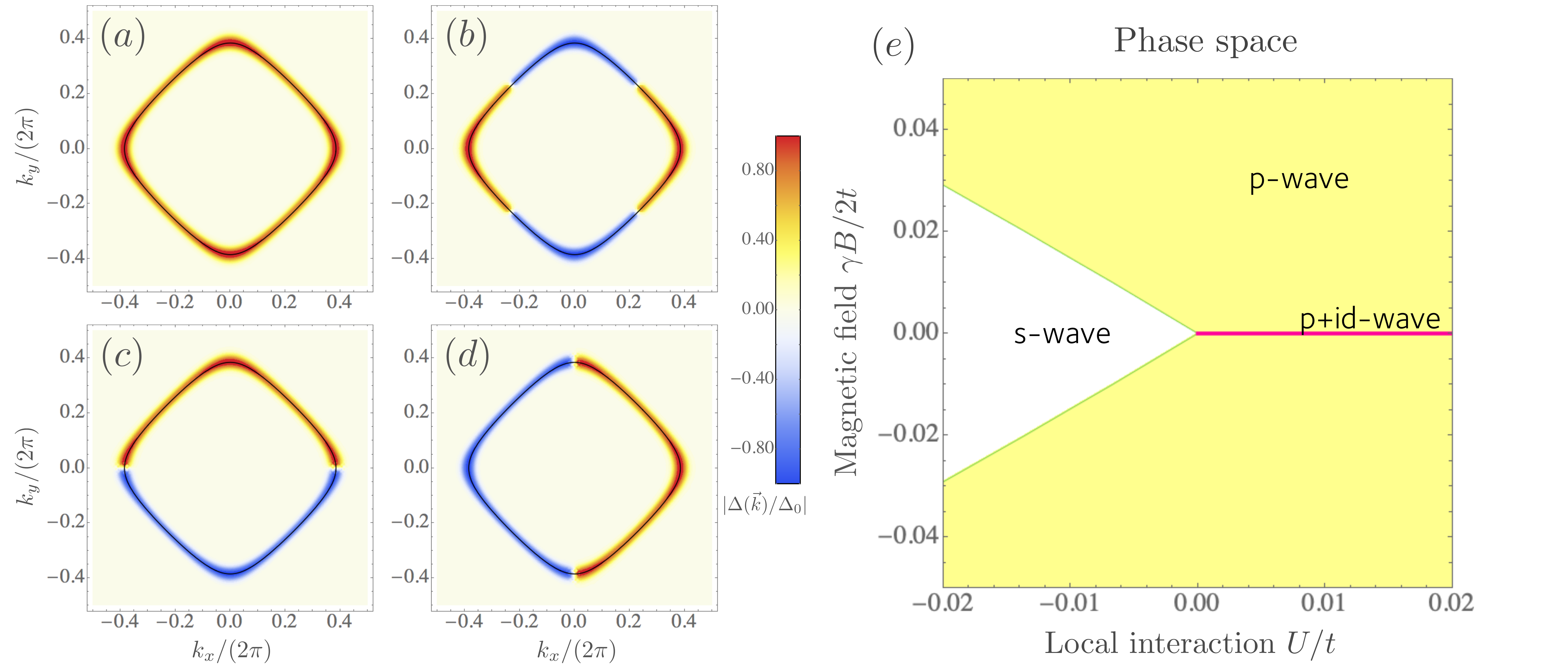}
\caption{
(a) - (d) Solutions of the gap equation~(\ref{eq.gap}) in a two-dimensional fermion gas. The initial state for the iterative solution is chosen with (a) $s$-wave, $\Delta_{\text{ini}} (\vec{k}) = t$, (b) $d$-wave, $\Delta_{\text{ini}} (\vec{k}) = t (\cos (k_x) - \cos (k_y) )$, and (c) and (d) $p$-wave symmetry, $\Delta_{\text{ini}} (\vec{k}) = t \sin (k_x)$ and $t \sin (k_y)$, respectively.
The other parameters are $k_c = 0.02 \pi$, $V_0 = - 0.1 t $, $U = 0$ and $\mu = - 0.5 t$.
(e) Dominating pairing symmetry vs. the magnetic field energy $\gamma B$ and the local Feshbach interaction $U$. The cavity-mediated interaction strength is $V_0 = - 0.1 t$, and the chemical potential $\mu = -3 t$.
}
\label{fig.phase-space}
\end{figure*}

Specifically, we consider a fermionic atom gas loaded into a two-dimensional optical square lattice potential with lattice period $a$. Its Hamiltonian is given by
$H_0 = - t \sum_{\langle \vec{n}, \vec{m} \rangle, \sigma} c^{\dagger}_{\vec{n}, \sigma} c_{\vec{m}, \sigma} - \mu \sum_{\vec{n}} n_{\vec{n}}$, 
where $t$ is the hopping integral, and $\mu$ the chemical potential. The summation $\langle \vec{n}, \vec{m} \rangle$ indicates summation over adjacent sites $\vec{n}$ and $\vec{m}$.
Additionally, we consider a contact interaction which we will describe later in the manuscript (see Eq.~(\ref{eq.H_U})). 
The lattice is placed into a cavity setup sketched in Fig.~\ref{fig.setup}, where two cavities oriented along the optical lattice axes induce atomic interactions as described theoretically in detail, e.g., in \cite{Camacho17}. 

The atoms are driven off-resonantly by a pump laser, and can scatter energy into ring cavity fields oriented along the $x$- and $y$-direction of the optical lattice, thereby experiencing a momentum kick $\pm k_c$ in the direction of the cavity fields. Due to energy conservation, this process can only take place, when the cavity photon is reabsorbed by another atom and re-emitted into the laser. Adiabatically eliminating the cavity fields and excited atomic levels, this creates an effective atomic interaction (see Supplemental Material (SM))
\begin{align}
H_{\text{cavity}} &= \frac{V_0}{2 N}   \sum_{\vec{k}, \vec{k}', \vec{k}_c, \sigma, \sigma'} c^{\dagger}_{\vec{k} \pm \vec{k}_c, \sigma} c_{\vec{k}, \sigma} c^{\dagger}_{\vec{k}' \mp \vec{k}_c, \sigma'} c_{\vec{k}', \sigma'}, \label{eq.H_cavity}
\end{align}
where $N$ is the number of lattice sites, and $\vec{k}_c = k_c \hat{e}_i$ with $i = x,y$.
$V_0$ is the interaction strength which is proportional to the pump laser intensity. Its sign can be controlled by the sign of the detuning between the cavity resonance and the pump field frequency, and here we only consider $V_0 < 0$. 

In deriving Eq.~(\ref{eq.H_cavity}), we assume driving by a linearly polarised pump laser, such that the interaction affects both spin species identically.
As a consequence, the interaction can mediate pairing of both singlet and triplet pairs. 
Previous theoretical works describing cavity-mediated interactions of fermionic atoms had focused on one-dimensional lattices \cite{Colella18, Santiago16, Santiago16b, Camacho17}. In \cite{Colella18}, the spin states were energetically separated, such that the cavity coupling was shown to induce either singlet superfluidity or spin-density order. 
Camacho-Guardian et al.~\cite{Camacho17} investigate the same effective interaction mechanism as considered here. Yet their focus lies on a situation, where the cavity has similar frequency as the standing laser wave creating the optical potential, such that $|\vec{k}_c| \sim \pi$ (where we measure $k$ in units of $1 / a$). This regime is signified by the competition between superfluid pairing and various charge or spin density wave instabilities.
Here, we focus instead on a regime with $|\vec{k}_c| \ll \pi$, where we show that different superfluid instabilities compete. This regime could be realised either by employing an infrared or a terahertz cavity, or by tilting the axis of an optical cavity with respect to the atomic gas, such that only the cavity wavevector's projection onto the gas plane is transferred to the atoms \cite{Camacho17}. In this regime, there is no nesting of wavevectors, and competing density wave instabilities are suppressed. 

We restrict our attention to situations where the induced atomic interactions are weaker than the kinetic energy scale, and only consider the atomic interaction in the Cooper channelgiven by 
$V_{\text{Cooper}} = \frac{1}{2 N} \sum_{\vec{k}, \vec{k}', \sigma, \sigma'} V_{\vec{k}, \vec{k}'} c^{\dagger}_{\vec{k}, \sigma} c^{\dagger}_{-\vec{k}, \sigma'} c_{- \vec{k}', \sigma'} c_{\vec{k}, \sigma}$, with $V_{\vec{k}, \vec{k}'} = V_0 \sum_{\vec{k}_c} [ \delta (\vec{k}- \vec{k}' - \vec{k}_c) + \delta (\vec{k}- \vec{k}' + \vec{k}_c) ]$. In the following, we will investigate the pairing symmetry using the weak-coupling BCS mean field approach. Since the long-range (in fact, in our setup it is infinite-range) nature of the cavity-mediated interaction implies that any atom will always interact with a large number of other atoms, this approach is well justified. This was found to be the case even in one dimension in \cite{Colella18} using bosonisation methods.

We decouple the electron interactions using the mean fields $\Delta_{\sigma \sigma'} (\vec{k}) = N^{-1} \sum_{k'} V_{k k'} \langle c_{- \vec{k'}, \sigma'} c_{\vec{k'}, \sigma} \rangle$. Going to the continuous limit, it is straightforward to numerically solve the resulting gap equation at zero temperature by iteration,
\begin{align}
\Delta (\vec{k}) &= - \int_{BZ} \frac{d^2 k'}{ (2\pi)^2 } \frac{V_{\vec{k}, \vec{k}'}}{2} \frac{\Delta (\vec{k}')}{ \sqrt{ \epsilon_{\vec{k}'}^2 + | \Delta (\vec{k}') |^2}  }, \label{eq.gap}
\end{align}
where $\epsilon_{\vec{k}} = - 2t \sum_{i = x, y} \cos (k_i)  - \mu$ denotes the normal state energy dispersion. 

Depending on how we choose the initial state for the iteration, we can realise different pairing symmetries. In particular, as shown in Fig.~\ref{fig.phase-space}(a)-(d), we find that $s$-, $p$- and $d$-wave symmetries yield almost identical mean field strengths, largely regardless of chemical potential or interaction strength. 
We quantify the pairing strength by the maximal mean field value $\Delta_0$ in the simulations shown in Fig.~\ref{fig.phase-space}. As a check, we also evaluated the integrated strength $\int_{BZ} \vert \Delta (\vec{k}) \vert^2 d^2 / (2\pi)^2$, yielding the same degeneracy. 
The top row symmetries correspond to singlet pairing with $s$-wave [in Fig.~\ref{fig.phase-space}(a)] or $d$-wave character [in Fig.~\ref{fig.phase-space}(b)], while the bottom row corresponds to triplet pairing, where the mean fields have nodes at either $k_y = 0$ [in Fig.~\ref{fig.phase-space}(c)] or $k_x = 0$ [in Fig.~\ref{fig.phase-space}(d)]. Other pairing symmetries with a larger number of nodes result in weaker mean field strengths. 
To further check that these three symmetries are the only dominating ones, we have also analysed the leading eigenvalues of the linearised gap equation on the Fermi surface (see SM) \cite{Romer15}. 

The observed three-fold degeneracy is a direct consequence of the lattice symmetry, and the interaction~(\ref{eq.H_cavity}) with equal strength along both lattice axes. 
The interaction is restricted along the cavity axes [see Eq.~(\ref{eq.H_cavity})], and therefore symmetries which take their maxima along these axes such as $d_{x^2-y^2}$ are selected. Orienting the cavities along the anti-diagonals, for instance, would favour $d_{xy}$-pairing instead.
If we increase the cavity wavevector $k_c$ or reduce the interaction strength $V_0$, we find that the mean field solutions develop a more complicated structure that covers larger areas on the Brillouin zone (see SM), rather than being strongly localised around the Fermi surface as depicted in Fig.~\ref{fig.phase-space}. 
In this manuscript, we focus on the low-$k_c$ regime, where the mean fields peak at the Fermi surface.

Introducing perturbations will break the degeneracy between the symmetries, and determine the realised superfluid state.
Here, we focus on the simultaneous application of two such perturbations: First, we include a local contact interaction,
\begin{align}
H_U &= U \sum_{\vec{m}} n_{\vec{m}, \uparrow}  n_{\vec{m}, \downarrow}, \label{eq.H_U}
\end{align}
where $U$ denotes the Feshbach-induced local atomic interaction, which can be chosen either attractive ($U < 0$) or repulsive ($U > 0$). It is included in the gap equation~(\ref{eq.gap}) by replacing $V_{\vec{k}, \vec{k}'} \rightarrow V_{\vec{k}, \vec{k}'} + U$. Second, we consider a weak magnetic field along the $z$-direction,
\begin{align}
H_B &= - \frac{\gamma B}{2} \sum_{\vec{m}} \left( n_{\vec{m}, \uparrow} - n_{\vec{m}, \downarrow} \right),
\end{align}
which lifts the spin degeneracy. It splits the Fermi surface into different spin sectors, and thus creates a spin imbalance.

\begin{figure}[t]
\centering
\includegraphics[width=0.49\textwidth]{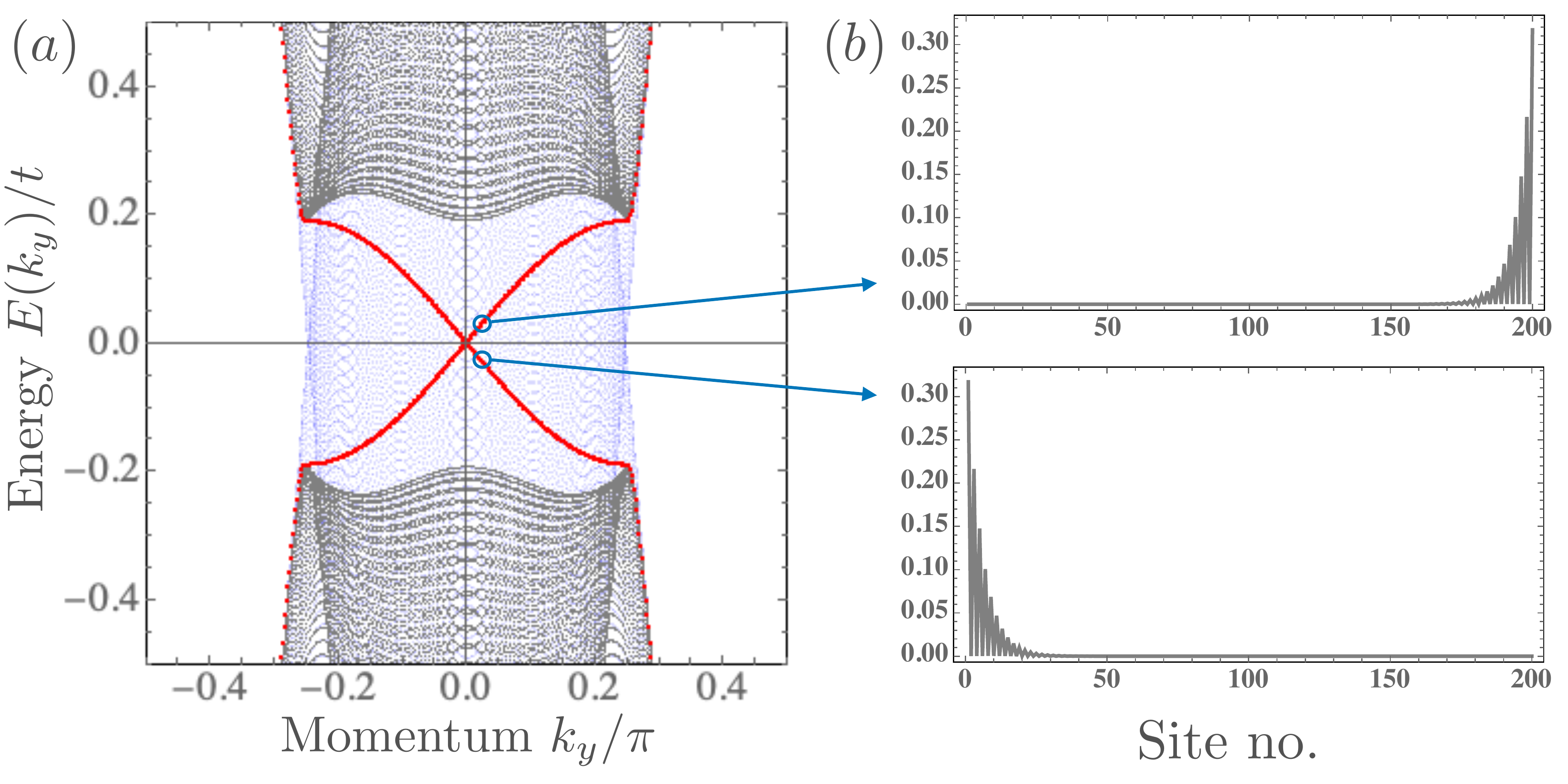}
\caption{
(a) Section of the band structure of a finite strip with $n = 200$ sites in width and periodic boundary conditions in $y$-direction, featuring $p$-wave pairing in the dominant spin-up channel. The spin-down states are indicated by semi-transparent blue dots. The parameters are $\mu = - 2 t$, $V_0 = - 0.1 t$, and magnetic field $\mu_0 B = 0.02 t$.
(b) Density plots of the topologically protected edge states, $\vert \psi_{\text{edge}} \vert^2$, at each side of the strip. 
}
\label{fig.p-wave}
\end{figure}

Fig.~\ref{fig.phase-space}(e) shows the dominating pairing symmetry as a function of the local interaction $U$ and the magnetic field $B$. We find that a local attractive interaction ($U < 0$) enhances the $s$-wave pairing, and pushes the system into a topologically trivial state. Conversely, a local repulsive interaction destroys the $s$-wave. The other symmetries' wavefunctions all feature a node at the origin, and are thus not affected by the local interaction.

When, in addition, the magnetic field strength $B$ is increased, the $p$-wave pairing becomes dominant. Small magnetic fields do not affect the singlet pairing, but change the chemical potential of the triplet amplitude (see SM).
As a consequence, the majority spin pairing amplitude becomes dominant over the singlet pairing.
The magnetic field strength at which the crossover to $p$-wave pairing takes place depends on the chemical potential, and consequently so does the critical magnetic field strength, at which this crossover takes place. In Fig.~\ref{fig.phase-space}(e), we chose a fairly low filling factor, such that a weak field is sufficient to favour $p$-wave pairing. 

Fig.~\ref{fig.p-wave} depicts the band structure of a quasiparticle Hamiltonian with $p$-wave pairing in the majority spin sector, by diagonalising the Hamiltonian on a finite strip.  
Each edge features a band of edge states with positive dispersion on the right boundary, and a negative dispersion curve on the other one. These states are covered by non-condensed atomic  states in the minority spin channel.
Since the free energy is typically minimised in a $J = 0$ state \cite{VollhardtWoelfle90}, the sign of the magnetic field - which determines the majority spin orientation - also determines the chirality of the condensate, and therefore the dispersion of the edge states.

\begin{figure}[t]
\centering
\includegraphics[width=0.49\textwidth]{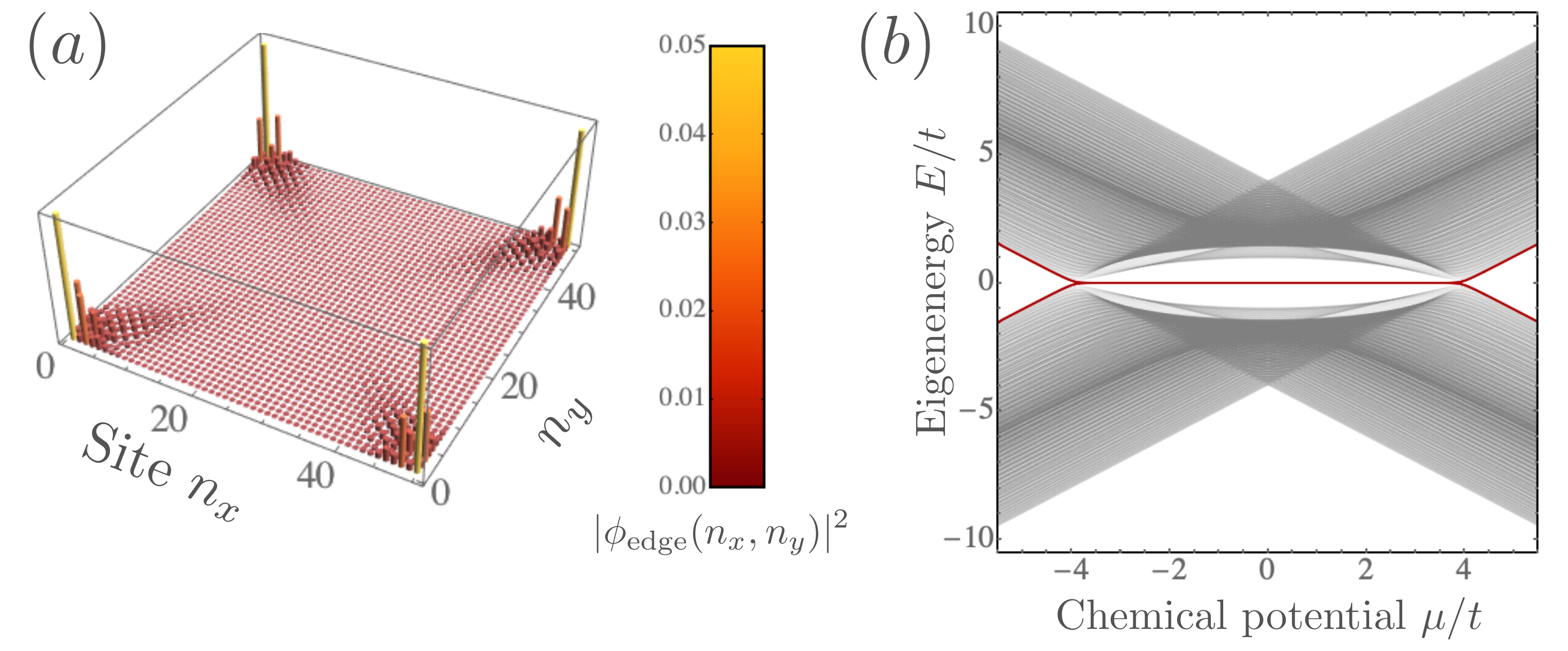}
\caption{
(a) The density of localised Majorana bound state wavefunctions $|\phi_{\text{edge}} (n_x, n_y) |^2$ of the $p + i d$-state, corresponding to a local repulsive interaction on the right side of panel (a), are plotted on a lattice with $50 \times 50$ sites. The mean field strength is fixed at $\Delta_0 = 0.19 t$, and $\mu = - 0.5 t$.
(b) The eigenenergies of a $30 \times 30$ site square lattice Hamiltonian in the $p+i d$-state are shown vs. the chemical potential $\mu$. The mean field strength is enhanced to $\Delta_0 = t$ to improve the visibility of the gap. 
}
\label{fig.2D}
\end{figure}

We next consider the case $B = 0$ specifically.
In the absence of a magnetic field, a local repulsive interaction will suppress the $s$-wave, but it does not affect the $p$- and $d$-wave symmetries.
We investigate the competition between the $p$- and $d$-wave phases by adopting a standard mean field Hamiltonian for unconventional superconductivity that captures the symmetry properties of the leading eigenfunctions. 
The relative phase between the $p$- and $d$-wave mean field components cannot be deduced from our mean field solutions, so we obtain it by numerically minimising the ground state energy of the system (see SM), and find the phase to be $ \pm \pi /2 $. This means the system forms a nodeless, helical superfluid \cite{Kallin16}. 
In contrast to most common models of chiral superconductivity in condensed matter, where chirality is created by strong spin-orbit coupling or by externally breaking time-reversal symmetry (as in Fig.~\ref{fig.p-wave}), here we find it to be the consequence of an "accidental" degeneracy of two competing phases, creating a frustrated system. It is not enforced by a symmetry of the system, but rather by the infinite range of the cavity-mediated interaction: a node (as in $p$- or $d$- wave pairing) in the two-atom wavefunction only results in a negligibly small energy penalty, and consequently the two mean fields are almost perfectly degenerate.   
In this situation, in order for the system to avoid nodes in the emerging band gap, the Cooper pairs acquire a chiral character, and time reversal symmetry is spontaneously broken. Experimentally, the chirality of the condensate could be tested through the measurement of density-spin or of density-current correlations \cite{Mraz05}. Topological order could be detected in time-of-flight measurements as described in \cite{Alba11}.

Topological properties of this $p+id$ chiral state were recently investigated in \cite{Wang18}. It was shown to be a \textit{second-order} topological state \cite{Benalcazar17, Schindler18} that features zero-dimensional edge states which are localised in the corners of the system. This behaviour can be understood when starting from a $p$-wave superfluid, which features a one-dimensional edge state dispersion (see Fig.~\ref{fig.p-wave}). 
The admixing of the complex $d$-wave order parameter gaps out the edge state dispersion of the $p$-wave state, leaving behind only localised Majorana states at the sample corners, where the $d$-wave mean field vanishes.
In Fig.~\ref{fig.2D}(a), these corner states are shown on a finite lattice with $p + i d$ pairing. Due to their topological protection, these states are fixed at zero energy regardless of the mean field strength or local perturbations of the Hamiltonian. We plot the variation of the eigenenergies with the chemical potential in Fig.~\ref{fig.2D}(b), certifying that the topological corner states are fixed at zero energy for any $\mu \in ( - 4t, 4t )$. At $| \mu | = 4 t$, the gap closes, and the system undergoes a topological phase transition to a trivial state. 

Finally, we remark that a conceptually similar situation was investigated in \cite{Wang17}, where degenerate pairing instabilities to $s$- and $p$-waves were considered in a three-dimensional condensed-matter system. In the presence of time-reversal symmetry, it was found that the phase transition between the two symmetries proceeds through an intermediate $s+ip$-state, which can be called an ``axion superconductor". Though we did not investigate this possibility here, the axion mode emerges from relative phase fluctuations of the symmetric singlet and the anti-symmetric triplet components, and could also be present in the $p + i d $-state we found here.

To conclude, we have found exotic $p+id$-states featuring Majorana fermions arising from the competition between cavity-induced, long-range and Feshbach-induced, local atom-atom interactions.
This proposed setup enables the continuous variation of the perturbations, and thus to scan the system through various topological phase transitions - between trivial and first- or second-order topological states, or between two distinct topological phases. 
Our approach further provides a new approach to create Majorana fermions in ultracold atoms, and could lead to an alternative platform for topological quantum computing \cite{Laflamme14}.
Rather than fine-tuning a microscopic Hamiltonian, it relies upon the controlled breaking of degeneracies of different pairing instabilities.
In the future, the insertion of local impurity atoms as quantum probes \cite{Streif16, Thomson16, Usui2018} could allow for the read-out and possibly the controlled manipulation of the localised Majorana modes created in such a setup.
Furthermore, with cavity design of materials also being discussed in condensed matter \cite{Sentef18, Schlawin18, Curtis18, Kiffner18, Allocca18, Bartolo18, Paracivini19, Mazza19}, ultracold atoms could form the an ideal platform to test these theoretical proposals in well-controlled settings. 

\section{acknowledgements}
\begin{acknowledgements} 
The research leading to these results has received funding from the European Research Council under the European Union's Seventh Framework Programme (FP7/2007-2013) Grant Agreement No. 319286 Q-MAC. D. J. acknowledges funding from the EPSRC grant EP/P009565/1 DESOEQ.
\end{acknowledgements}

\end{document}